\documentstyle[prb,preprint,aps]{revtex}
\begin{document}
\tightenlines
\draft
\title{Nuclear relaxation in the spin-1/2 antiferromagnetic chain
compound Sr$_2$CuO$_3$ --- comparison between theories and experiments}
\author{M. Takigawa$^1$\cite{byline1}, O. A. Starykh$^2$,
A. W. Sandvik$^3$, and R. R. P. Singh$^4$}
\address{$^1$IBM Thomas J. Watson Research Center, P.O. Box 218,
Yorktown Heights, NY 10598}
\address{$^2$Department of Physics, University of Florida, P. O. Box
118440, Gainesville, FL 32611}
\address{$^3$Department of Physics, University of Illinois at
Urbana-Champaign, Urbana, IL 61801}
\address{$^4$Department of Physics, University of California, Davis,
CA 95616}
\date{\today}
\maketitle
\begin{abstract}
The NMR relaxation data on Sr$_2$CuO$_3$ [Phys. Rev. Lett. {\bf 76}, 4612
(1996)] are reexamined and compared with the analytic theory of the
dynamic susceptibility in the S=1/2 antiferromagnetic Heisenberg chain
including multiplicative logarithmic corrections [Phys. Rev. Lett.
{\bf 78}, 539 (1997)].  Comparisons of the spin-lattice and the gaussian
spin-echo decay rates ($1/T_1$ and $1/T_{2G}$) and their ratio all show
good quantitative agreement.  Our results demonstrate the importance of
the logarithmic corrections in the analysis of experimental data for
quasi-1D systems and indicate that the dynamics of Sr$_2$CuO$_3$ is well
described by a S=1/2 one-dimensional Heisenberg model with a nearest
neighbor exchange.
\end{abstract}

\narrowtext

The spin-1/2 one-dimensional antiferromagnetic Heisenberg model
\begin{equation}
{\cal H} = J \sum_i {\bf S}_i \cdot {\bf S}_{i+1}
\end{equation}
continues to reveal novel many-body quantum effects
\cite{eggert941,eggert951}
in spite of its simplicity and extensive studies over many decades.
In particular, a good understanding of the dynamic properties at finite
temperature has been obtained only recently, using field theories
\cite{shulz861,eggert941,sach941,starykh971} and advanced numerical
techniques.\cite{starykh972} Experimentally, there are generally two
complementary methods to observe spin dynamics. The spin correlation function
over a wide frequency ($\omega$) and momentum ($q$) range can be measured by
neutron scattering experiments. \cite{tennant931} The nuclear spin-lattice
and the gaussian spin-echo decay rates ($1/T_1$ and $1/T_{2G}$) measured
by nuclear magnetic resonance (NMR) experiments \cite{taki961} provide
accurate information on $q$-averaged low frequency dynamic and static
susceptibilities, respectively.  Since it is the static and low frequency
($\hbar \omega \ll J$) properties at low temperatures ($T \ll J$) that
can be treated most accurately by field theories, NMR is particularly
suitable to test them.

A remarkable consequence of the field theories is the quantum critical scaling
behavior, \cite{shulz861} i.e, scaling of the dynamic susceptibility
$\chi (q, \omega , T)$ in the variables $c(q-\pi )/T$ and $\hbar \omega /T$
($c=\pi J/2$ is the spinon velocity). This implies a divergence of the
antiferromagnetic correlation length and the characteristic time scale of
spin fluctuations as $1/T$ at low temperatures and reflects the critical
nature of the ground state.  The scaling determines the temperature
dependences of the NMR relaxation rates as $1/T_1 = const$ and $1/T_{2G}
\propto 1/\sqrt T$. \cite{sach941} This behavior has indeed been observed
in the experiments on Sr$_2$CuO$_3$,\cite{taki961} which has the most
ideal one dimensional character known to date.\cite{keren931,ami951,moto961}
The theory also shows that the scaling is correct only approximately and
there is a multiplicative correction with logarithmic temperature
dependence.\cite{singh891,affleck891} However, including this correction to
the highest order actually degrades the agreement between theory and the
data on Sr$_2$CuO$_3$.\cite{taki961}

Very recently, an improved analytic theory of the logarithmic corrections
has been proposed by Starykh, Singh, and Sandvik, \cite{starykh971}
generalizing the scaling ansatz and the conformal mapping to include the
logarithmic factors.  In this paper, we reexamine the NMR data on
Sr$_2$CuO$_3$ reported in Ref.~\onlinecite{taki961} and compare them with
this theory.
In the following, we first describe the theoretical results for the NMR
relaxation rates both with and without the logarithmic corrections.
We then reanalyze the spin-echo decay data to extract $1/T_{2G}$ more
accurately, taking the nuclear spin fluctuations due to spin-lattice
relaxation into account, and also examine the appropriate value of $J$
in Sr$_2$CuO$_3$.  From the comparison, we found that including the
logarithmic corrections considerably improves the quantitative agreement.

Let us start with the expression for $1/T_1$ \cite{moriya561}
and $1/T_{2G}$ \cite{penn91,moriya621,taki941} due to the
magnetic hyperfine interaction
\begin{equation}
{\cal H}_{\rm hf} = \sum_{\alpha ,i,j}
A_{\alpha}^{ij} I_{i \alpha} S_{j \alpha} \; (\alpha = a, b, {\rm and}
\: c)
\end{equation}
between a nuclear spin at $i$ site and an electron spin at $j$ site,
\begin{mathletters}
\begin{equation}
{1\over T_1} = {k_B T\over \hbar^2} \int {dq\over 2\pi} \left\{
A_a ^2 (q) + A_b ^2 (q) \right\} {{\rm Im}\chi (q , \omega_0 )\over
\omega_0}
\end{equation}
\begin{equation}
\left( {1\over T_{2G}} \right) ^2 = {p\over 8\hbar^2} \left[ \int {dq\over
2\pi} A_c ^4 (q) \chi^2 (q) - \left\{ \int {dq\over 2\pi} A_c ^2 (q)
\chi (q) \right\} ^2 \right]  .
\end{equation}
\end{mathletters}
Here $\chi (q, \omega )$ ($\chi (q)$) is the dynamic (static)
susceptibility per spin in units of $(g\mu_B)^2$, $\omega_0$ is the NMR
frequency, and $A_{\alpha}(q)=\sum_j A_{\alpha} ^{ij} \exp (iqr_{ij})$.
The magnetic field is assumed to be applied along the $c$ direction.
The prefactor in Eq. (3b) is chosen for nuclei with spin 3/2 and is
valid when the spin-echo decay is measured on one of three NMR lines
split by the quadrupole interaction.  The isotopic abundance $p$ is
0.69 for $^{63}$Cu nuclei.  Since $\chi (q)$ and Im$\chi (q, \omega)$
are strongly peaked at $q=\pi$ (the antiferromagnetic wave vector),
$A_{\alpha}(q)$ in the integral can be replaced by $A_{\alpha}(\pi)$.

An analytic result for $\chi (q, \omega )$ at finite temperatures was
first obtained by Schultz, \cite{shulz861} using the bosonization
technique to transform the Heisenberg model to a free boson Hamiltonian.
The result satisfies the quantum critical scaling,
\begin{equation}
\chi (q, \omega ) = D f_1 (\tilde{q}, \tilde{\omega})/k_BT   ,
\end{equation}
where $\tilde{q}=c(q-\pi)/k_BT$, $\tilde{\omega}=\hbar\omega /k_BT$,
and $D$ is an unknown constant determining the overall magnitude of
$\chi (q, \omega )$.
It immediately follows from Eqs. (3) and (4) that $1/T_1 = const.$ and
$1/T_{2G} \propto 1/\sqrt{T}$.  In order to eliminate dependence on
material parameters, we define the following normalized dimensionless
NMR rates, which should be universal functions of $T/J$ and can be
compared directly with theories,
\begin{mathletters}
\begin{equation}
\left(1/T_1\right)_{\rm norm} = {2 \hbar J \over T_1 \left\{ A_x^2(\pi) +
A_y^2(\pi) \right\}}
\end{equation}
\begin{equation}
\left(\sqrt{T}/T_{2G}\right)_{\rm norm} =
\left({k_BT \over pJ} \right)^{1/2} {\hbar J \over A_z^2(\pi) T_{2G}}  .
\end{equation}
\end{mathletters}
Sachdev used the result by Schulz to calculate the NMR rates and
obtained \cite{sach941}
\begin{equation}
\left(1/T_1\right)_{\rm norm} = 2 D   ,
\left(\sqrt{T}/T_{2G}\right)_{\rm norm} = 1.1908 D ,
\end{equation}
with their ratio being a universal number
$(T_{2G}/T_1\sqrt{T})_{\rm norm} = 1.680$, which can be tested
experimentally. The $T$-dependence of $1/T_1$ and $1/T_{2G}$ and the
value of $T_{2G}/(\sqrt{T}T_1)$ indeed agree quite
well with the experimental data on Sr$_2$CuO$_3$. \cite{taki961}

However, the free boson theory neglects the marginally
irrelevant operator in the original Heisenberg Hamiltonian,
describing umklapp scattering processes,
that leads to a multiplicative logarithmic correction to $\chi (q,
\omega)$.  Sachdev has shown that to the leading order both
$(1/T_1)_{\rm norm}$ and $(\sqrt{T}/T_{2G})_{\rm norm}$ acquire an
identical multiplicative factor of $\ln ^{1/2}(T_0/T)$, where $T_0$
is the high energy cutoff of the order of $J$. \cite{sach941} However,
the agreement with the experimental data becomes worse if the factor
$\ln^{1/2}(J/T)$ is included.\cite{taki961}
Recently Starykh, Singh, and Sandvik have proposed a more
elaborate theory of the logarithmic corrections \cite{starykh971}
by adopting a simple ansatz generalizing the finite-size
scaling and the conformal mapping to correlation functions with
multiplicative logarithmic factors.  They showed that the susceptibility
takes the form,
\begin{equation}
\chi (q, \omega ) = D \ln^{1/2}(T_0/T) f_2 (\tilde{q}, \tilde{\omega},
\Delta )/k_BT ,
\end{equation}
with $\Delta = (1/4)\{1-{1 \over 2\ln(T_0/T)}\}$.  In addition to the
multiplicative factor, the logarithmic $T$-dependence of $\Delta$ also
breaks the scaling. The equal-time real-space correlation function obtained
from this formula agrees very well with Quantum Monte Carlo (QMC)
result for $8 \le J/T \le 32$ if $T_0/J \approx 4.5$ and $D \approx 0.075$.
The theoretical expressions for the NMR rates are \cite{starykh971}
\begin{mathletters}
\begin{equation}
\left(1/T_1\right)_{\rm norm} = 2D\times 2^{5/2-2 \Delta} \sin (2\pi\Delta)
I_1(\Delta) \ln^{1/2}(T_0/T)/\pi^2
\end{equation}
\begin{equation}
\left(\sqrt{T}/T_{2G}\right)_{\rm norm} = D\times
 2^{-5/2+2\Delta} \sin(2\pi\Delta) \Gamma^2(1-2\Delta) I_2(\Delta)
\ln^{1/2}(T_0/T)/\pi^{3/2}
\end{equation}
\end{mathletters}
with $I_1(\Delta) = \int_0^{\infty}dx{x \over (\sinh x)^{4\Delta}}$ and
$I_2(\Delta) = 4\int_0^{\infty}dx \left| {\Gamma (\Delta - ix) \over
\Gamma (1-\Delta-x)} \right|^4$.

The above finite-$T$ QMC estimate of the scale factor $D$ compares well with
recent $T=0$ numerical calculations ($0.06789$ \cite{hallberg} and $0.065$
\cite{koma}), but certainly may contain some errors arising from
non-asymptotic contributions in the temperature regime studied. The
high-energy cut-off $T_0$ has also be determined from fits of QMC data
for the static staggered susceptibility and structure factor to their
corresponding analytic expressions.\cite{starykh972}  The results are
$T_{0,\chi}=3.9 \pm 0.3$ from the susceptibility data and
$T_{0,s}=5.1 \pm 0.2$ from the structure factor.
Within the accuracy of the procedure these numbers can be considered being
in good agreement with the value $T_0=4.5$ quoted above.

We now turn to the experimental results.  NMR experiments on $^{63}$Cu
nuclei in Sr$_2$CuO$_3$ have been performed by Takigawa {\it et al}..
\cite{taki961}
They have measured $1/T_1$ along the three crystal axes and $1/T_{2G}$
along the $c$ direction.  A static approximation was used to obtain
$1/T_{2G}$ from the spin-echo decay data, namely time dependence
of $I_z$ due to spin-lattice relaxation process was neglected.
Although this is a reasonable approximation in Sr$_2$CuO$_3$, for which
$(1/T_1)_c$ is more than an order of magnitude smaller than
$(1/T_{2G})_c$, it is desirable to take the $I_z$ fluctuation effects
into account to obtain more accurate values of $1/T_{2G}$.  Recently,
Curro {\it et al}. \cite{curro971} have derived an highly accurate
analytic expression of spin-echo decay in the presence of the
$I_z$ fluctuations, using the gaussian approximation proposed by Recchia
{\it et al}.. \cite{recchia961}    We fitted the
spin-echo decay data in Sr$_2$CuO$_3$ to this expression and extracted
$1/T_{2G}$ as defined in Eq. (3b).

In order to convert the measured NMR rates to the normalized units
defined in Eq. (5), we need the values of $A_{\alpha}(\pi)$ and $J$.  The
values of $A_{\alpha}(\pi)/\sqrt{J}$ were determined from the
width of the characteristic broad background of the NMR spectra due to a
field-induced staggered magnetization near impurities, \cite{taki971}
using the calculation of alternating local susceptibility near chain ends
by Eggert and Affleck. \cite{eggert951}  The accuracy of the these
values depends on how closely the actual impurities (most likely the
holes in the Zhang-Rice singlet states \cite{zhang881}
due to excess oxygen) behave as
chain ends.  We consider it to be of the order of 10\%.  The anisotropy
of $A_{\alpha}$ can be determined much more accurately, with an uncertainty
of a few percents.

The exchange $J$ was estimated to be $2200 \pm 200$ K \cite{moto961}
by fitting the temperature dependence of the uniform susceptibility
to the theory
by Eggert, Affleck, and Takahashi. \cite{eggert941}  This value was used
in the analysis of the previous NMR experiments. \cite{taki961,taki971}
Suzuura {\it et al}., on the other hand, measured the optical absorption
spectrum at $T$=32 K, \cite{suzu961} which was interpreted to be due to
simultaneous phonon and spin excitations.  Lorenzana and Eder
\cite{loren971} analyzed the spectrum and obtained $J$=2850 K.  Since
this value is directly determined from the sharp peak of the spectrum
located exactly at $\pi J/2$ (the top of the des Cloizeaux-Pearson mode)
except for a small shift corresponding to the optic phonon frequency
($\sim 0.08$ eV), this latter value should be more reliable.

We speculate that the discrepancy may be resolved if one takes into account
lattice degrees of freedom. A temperature dependence of $J$ naturally
results from the thermal lattice expansion. In fact, we recognize a slight
systematic deviation between the $\chi (T)$ data and the theoretical
curve assuming a constant $J$ (Fig. 4 in Ref. ~\onlinecite{moto961}).
Since the measured susceptibility includes the
constant orbital and diamagnetic contributions, which are not known well,
we can choose $J=2850$ K at $T=0$ and allow $J$ to change with temperature
to reproduce the $\chi (T)$ data. We found that the $\chi (T)$ data can be
explained if $J$ is reduced to 2530 K at $T=800$K, which is the highest
temperature of the measurements. Since $J \propto t^4$ and
$t \propto d^{-3.5}$, where $t$ is the Cu to O transfer integral and
$d$ is the Cu-O distance, \cite{harr801} thermal expansion of
0.8 \% over 800 K is enough to account for this change of $J$. This
seems quite plausible.

A recent QMC study of the Heisenberg chain including dynamic (fully quantum
mechanical) phonons indicates that the fluctuations in $J$ may furthermore
lead to an apparent shift of $J$ as obtained from $\chi (T)$. A fit to the
Heisenberg $\chi (T)$, assuming a constant $J$, gives a result which is
lower than the actual average spin-spin coupling in the spin-phonon model.
\cite{sandvik97} This effect may be the reason for the different values of
$J$ obtained from $\chi (T)$ and the optical absorption. We use $J=2850$ K
in the following analysis but show also the results for $J=2200$ K to
indicate how sensitively the results depend on the value of $J$.

Finally, we are at a position to compare the experimental results with
the theory.  Fig. 1 shows the results for $(1/T_1)_{\rm norm}$. Since
the ratio $A_{\alpha} (\pi)/\sqrt{J}$ is determined from experiments,
$(1/T_1)_{\rm norm}$ is independent of the choice of $J$.  The solid line
shows the result of Eq. (8a) with $T_0/J=4.5$ and $D=0.062$, while the
dotted line is the $\ln ^{1/2}(J/T)$ dependence with a magnitude chosen
arbitrarily.  It is clear that the theory by Starykh {\it et al}. shows
much weaker T-dependence than $\ln ^{1/2}(J/T)$ and is closer to the
experimental data.  The value $D=0.062$ agrees with the value $D=0.075$
determined from the comparison with the QMC results of the real-space
correlation function \cite{starykh971} within the uncertainty due to
errors in $A_{\alpha}(\pi)/\sqrt{J}$ and numerical determination of D.

Fig. 2 shows the results for $(\sqrt{T}/T_{2G})_{\rm norm}$.  The open
(filled) circles are the experimental results for $J=2850$ K ($J=2200$ K)
adjusted for the $I_z$ fluctuations. The crosses are the results
for $J=2850$ K in the static approximation.  The correction for the
$I_z$ fluctuations reduces the value of $1/T_{2G}$ by 6 - 12 \%.
The solid and the dotted lines are the theoretical results of Eq. (8b)
for $D=0.062$ and $D=0.07$, respectively ($T_0/J=4.5$).  We  remark
that only the first term in Eq. (3b), which we call the ``scaling part'',
can be calculated reliably by the field theory, giving the results of Eq.
(8b).  The second term in Eq. (3b) is the square of the local
susceptibility, which cannot be handled properly by continuum effective
field theory. \cite{starykh971}  It is smaller than the
scaling part by a factor $(T/J)\ln ^2 (T_0/T)$.  QMC calculations
\cite{sandvik95} are useful to examine the relative magnitude of the
second term.  The filled
and open triangles in Fig. 2 show the QMC results for the full and the
scaling part, respectively.  The QMC results for the scaling part agree
very well with the analytic results with $D=0.07$.  The analytic results
for $D=0.062$ (the best value to fit the $1/T_1$ data) are in good
agreement with the experimental results for $J=2850 K$ (filled circles)
if the negative contribution from the second term in Eq. (3b) (the
difference between the two sets of QMC results) is taken into account.

Fig. 3 shows the ratio of Eqs. (5a) and (5b).  We used the $1/T_1$ data
for $H \parallel a$ and $H \parallel b$, since the hyperfine form factor
$A_x^2(q) + A_y^2(q)$ for $H \parallel c$ is relatively small at
$q= \pi$.  This ratio provides a particularly stringent test for
theories, since uncertainties in the values of $D$ and $A_{\alpha}(\pi)$
cancels out and the only parameter that needs to be known is the exchange
$J$.  The results of the analytic theory are between the two sets of
experimental data for different values of $J$.  It predicts a weak
logarithmic $T$-dependence, consistent with the experimental data,
with an infinite slope at $T=0$.  The value at $T=0$ shown by the
square in Fig. 3 is equal to the scaling result 1.680, which is
temperature independent.  Once again, including the logarithmic
corrections improves the agreement with the experimental data.

In summary, $1/T_1$, $1/T_{2G}$, and their ratio in Sr$_2$CuO$_3$ all
show good quantitative agreement with the results of the analytic field
theory that takes the logarithmic corrections into account.  Our results
demonstrate the importance of the logarithmic corrections in analyzing
experimental data for quasi-1D systems, and that the dynamics in
Sr$_2$CuO$_3$ is well described by S=1/2 1D Heisenberg model with a
nearest neighbor exchange. We suggest that remaining small discrepancies,
such as the different estimates of $J$ from the magnetic susceptibility
and optical absorption experiments, may be due to spin-phonon
interactions.

We would like to thank S. Sachdev for interesting discussions and C. P.
Slichter for communication on the analysis of spin-echo decay.  This work
was supported by the National Science Foundation under Grants
DMR-89-20538 (AWS) and DMR-96-16574 (RRPS).

\begin{figure}
\caption{Results for $(1/T_1)_{\rm norm}$.  Symbols are the experimental
data for different field directions.  The solid and dotted
lines show the analytic theory Eq. (8a) with $D=0.062$ and the
$\ln ^{1/2} (J/T)$ dependence, respectively.}
\label{fig1}
\end{figure}

\begin{figure}
\caption{Results for $(\protect \sqrt{T}/T_{2G})_{\rm norm}$.  The filled
and open circles are the experimental data for different values of $J$
corrected for the $I_z$ fluctuations.  The crosses are the data for
$J=2850$ K without the corrections.  The filled and open triangles show
the QMC results for the full and the scaling (the first term in Eq. (3b))
parts, respectively.  The results of the analytic theory Eq. (8b) for the
scaling part is shown by the solid ($D=0.062$) and the dotted ($D=0.07$)
lines.}
\label{fig2}
\end{figure}

\begin{figure}
\caption{Results for the ratio of $(1/T_1)_{\rm norm}$ and
$(\protect
\sqrt{T}/T_{2G})_{\rm norm}$.  Symbols are the experimental data for
different values of J and field directions.  The result of the analytic
theory (the ratio of Eqs. (8a) and (8b)) is shown by the line.  Its
value at $T=0$ (square) coincides the scaling result.}
\label{fig3}
\end{figure}


\begin{references}
\bibitem[a]{byline1}
Permanent address: Institute for Solid State Physics, University of
Tokyo, Roppongi, Minato-ku, Tokyo 106, Japan.
\bibitem{eggert941} S. Eggert, I. Affleck, and M. Takahashi, Phys. Rev.
Lett. {\bf 73}, 332 (1994).
\bibitem{eggert951} S. Eggert and I. Affleck, Phys. Rev. Lett. {\bf 75},
934 (1995).
\bibitem{shulz861} H. J. Schulz, Phys. Rev. B{\bf 34}, 6372 (1986).
\bibitem{sach941} S. Sachdev, Phys. Rev. B{\bf 50}, 13006 (1994).
\bibitem{starykh971} O. A. Starykh, R. R. P. Singh, and A. W. Sandvik,
Phys. Rev. Lett. {\bf 78}, 539 (1997).
\bibitem{starykh972} O. A. Starykh. A. W. Sandvik, and R. R. P. Singh,
Phys. Rev. B{\bf 55}, 14953 (1997).
\bibitem{tennant931} D. A. Tennant {\it et al}., Phys. Rev. Lett.
{\bf 70}, 4003 (1993).
\bibitem{taki961} M. Takigawa {\it et al}., Phys. Rev. Lett. {\bf 76},
4612 (1996).
\bibitem{keren931} A. Keren {\it et al}., Phys. Rev. B{\bf 48}, 12926
(1993).
\bibitem{ami951} T. Ami {\it et al}., Phys. Rev. B{\bf 51}, 5994 (1995).
\bibitem{moto961} N. Motoyama, H. Eisaki, and S. Uchida, Phys. Rev. Lett.
{\bf 76}, 3212 (1996).
\bibitem{singh891} R. R. P. Singh, M. E. Fisher, and R. Shankar, Phys.
Rev. B{\bf 39}, 2562 (1989).
\bibitem{affleck891} I. Affleck {\it et al}., J. Phys. A{\bf 22}, 511
(1989).
\bibitem{moriya561} T. Moriya, Prog. Theor. Phys. {\bf 16}, 23, 641
(1956).
\bibitem{penn91} C. H. Pennington and C. P. Slichter, Phys. Rev. Lett.
{\bf 66}, 381 (1991).
\bibitem{moriya621} T. Moriya, Prog. Theor. Phys. {\bf 28}, 371 (1962).
\bibitem{taki941} M. Takigawa, Phys. Rev. B{\bf 49}, 4158 (1994).
\bibitem{hallberg} K. A. Hallberg, P. Horsch, and G. Martinez, \prb {\bf 52},
R719 (1995)
\bibitem{koma} T. Koma and N. Mizukoshi, J. Stat. Phys. {\bf 83}, 661 (1996)
\bibitem{curro971} N. J. Curro {\it et al}., Phys. Rev. B{\bf 56}, in
press.
\bibitem{recchia961} C. H. Recchia, K. Gorny, and C. H. Pennington,
Phys. Rev. B{\bf 54}, 4207 (1996).
\bibitem{taki971} M. Takigawa {\it et al}., Phys. Rev. B{\bf 55}, 14129
(1997).
\bibitem{zhang881} F. C. Zhang and T. M. Rice, Phys. Rev. B{\bf 37}, 3759
(1988).
\bibitem{suzu961} H. Suzuura {\it et al}., Phys. Rev. Lett. {\bf 76},
2579 (1996).
\bibitem{loren971} J. Lorenzana and R. Eder, Phys. Rev. B{\bf 55},
R3358 (1997).
\bibitem{harr801} W. A. Harrison, {\it Electronic Structure and Physical
Properties of Solids}, Freeman, San Francisco, (1980).
\bibitem{sandvik97} A. W. Sandvik, R. R. P. Singh, and D. K. Campbell,
preprint (cond-mat/9706046).
\bibitem{sandvik95} A. W. Sandvik, Phys. Rev. B {\bf 52}, R9831 (1995).

\end{references}
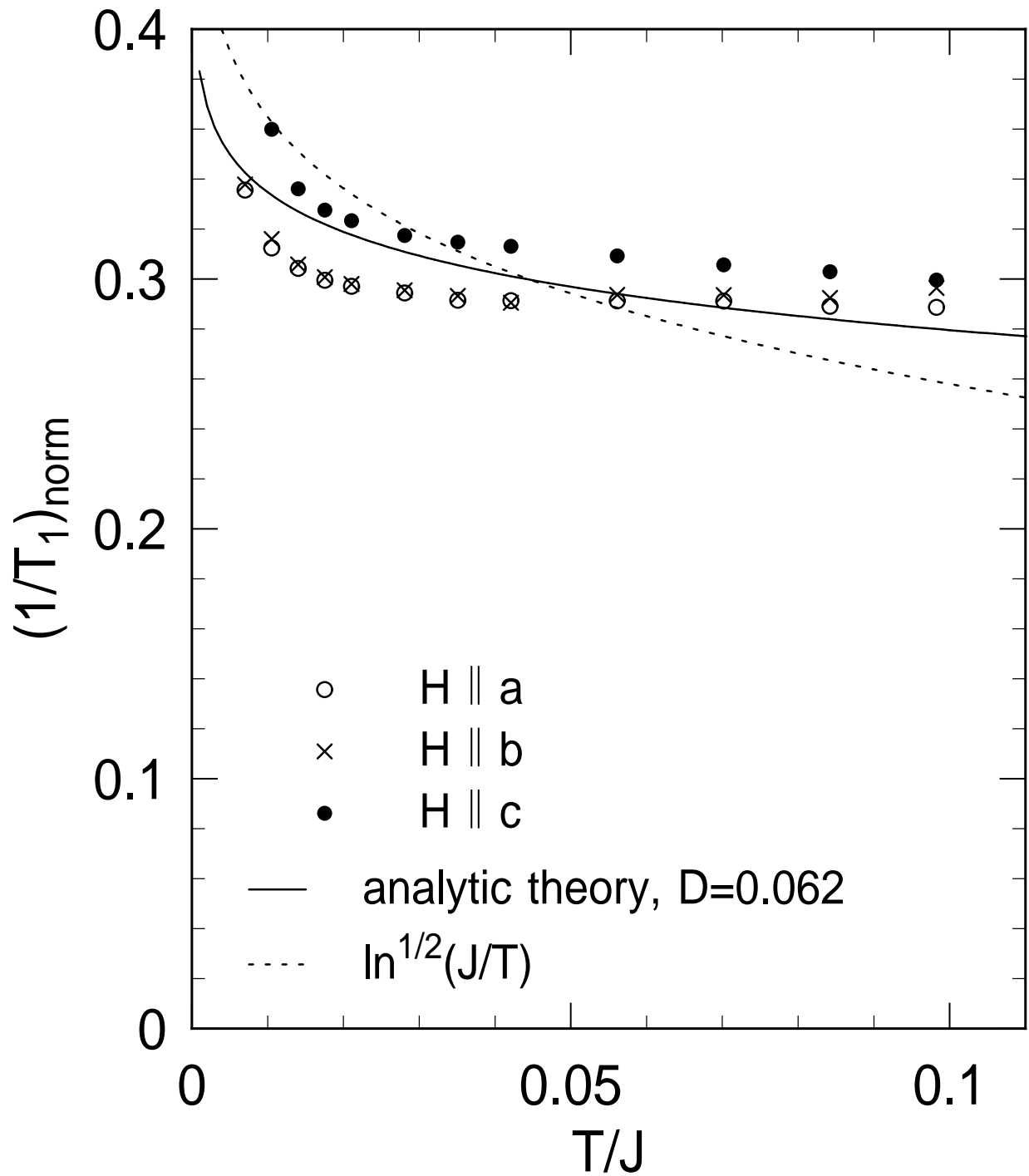
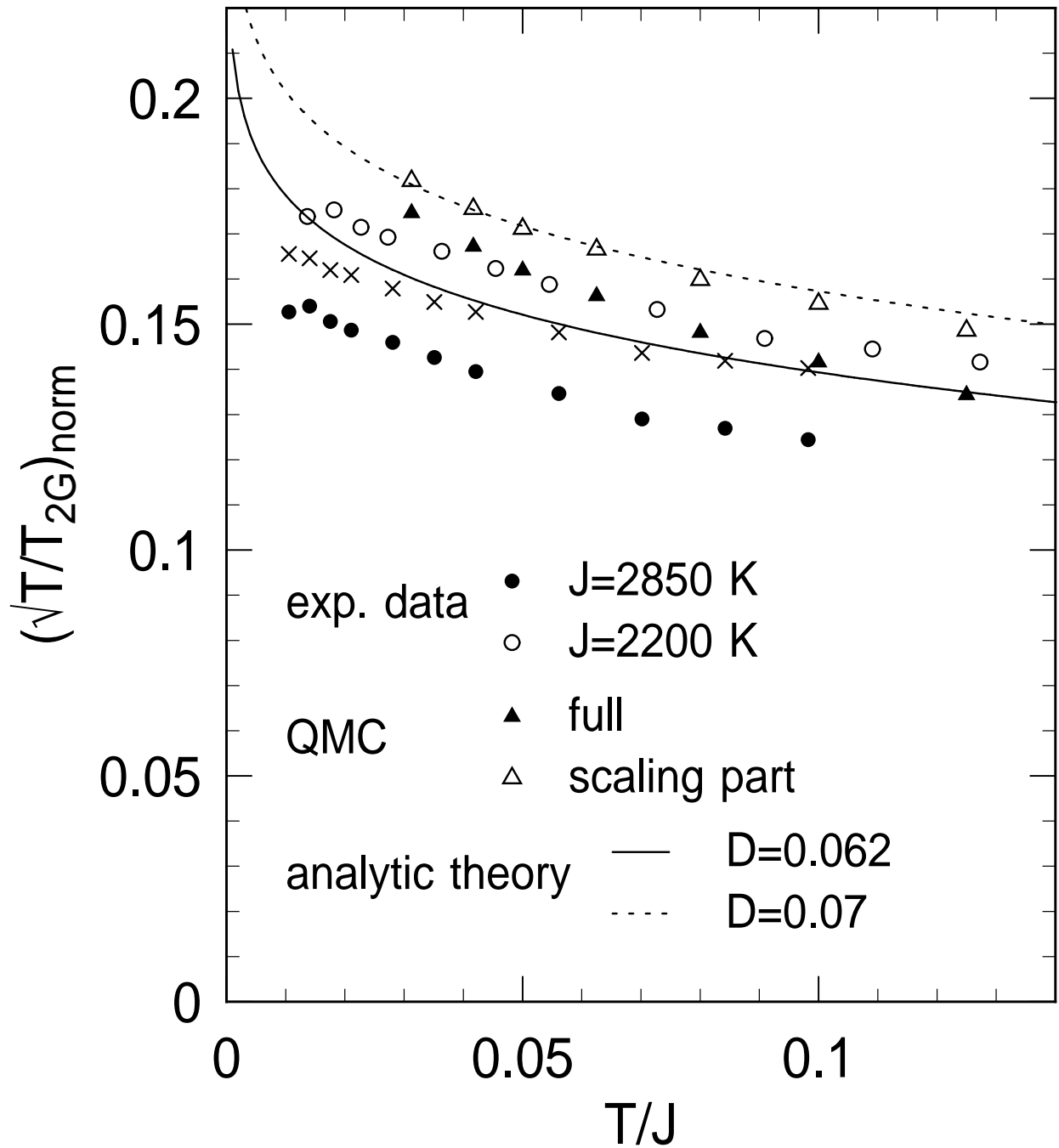
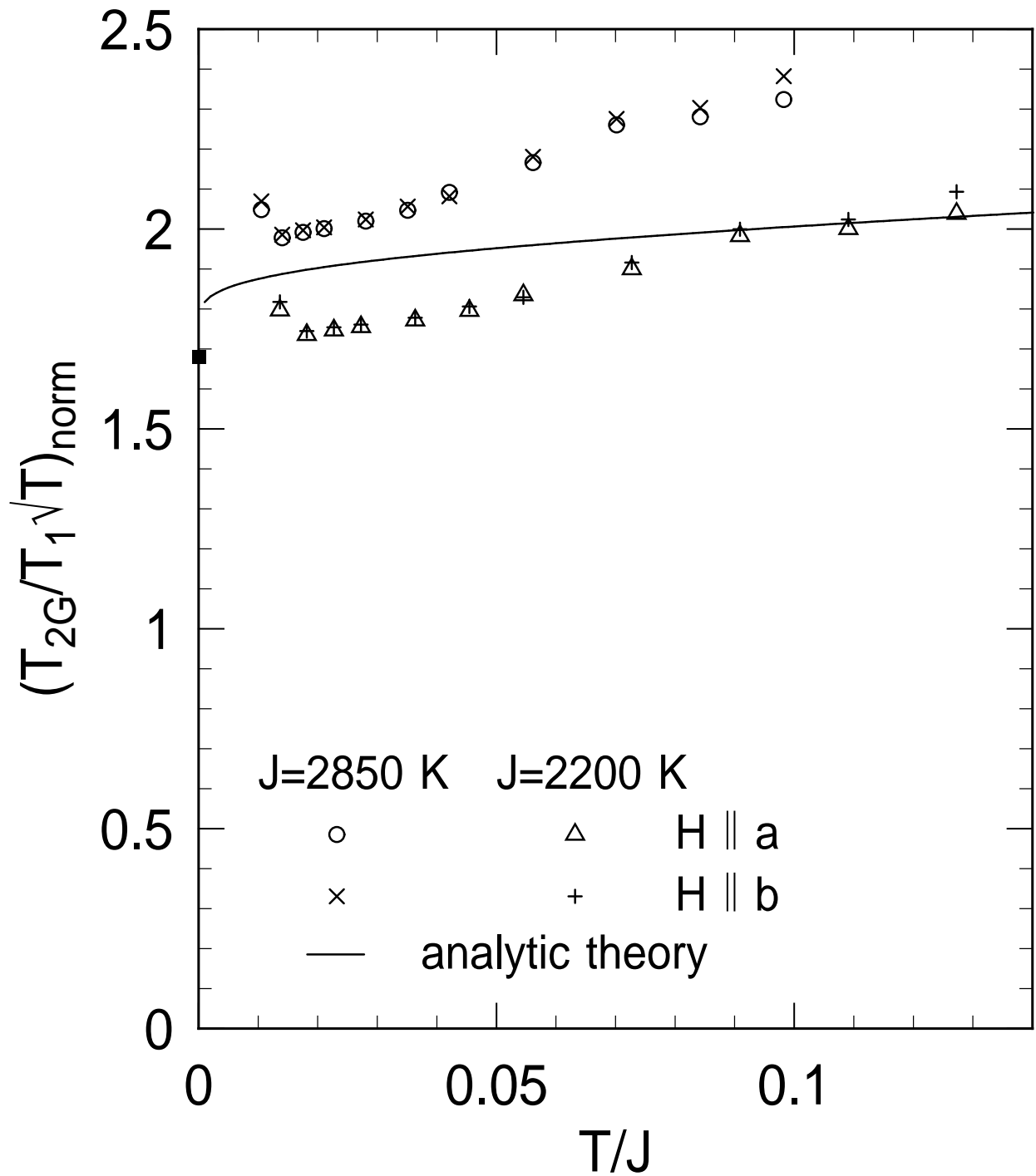
\end{document}